# Summarizing text to embed qualitative data into visualizations

Richard Brath*

Uncharted Software Inc.

**ABSTRACT**

Qualitative data can be conveyed with strings of text. Fitting longer text into visualizations requires a) space to place the text inside the visualization; and b) appropriate text to fit the space available. For quantitative visualizations, space is available in area marks; or within visualization layouts where the marks have an implied space (e.g. bar charts). For qualitative visualizations, space is defined in common text layouts such as prose paragraphs. To fit text within these layouts is a function for emerging NLP capabilities such as summarization.

**Index Terms**: K.6.1 [Management of Computing and Information Systems]: Project and People Management—Life Cycle; K.7.m [The Computing Profession]: Miscellaneous—Ethics

## 1 INTRODUCTION

Information visualization has tended to focus almost exclusively measurable data – quantitative, ordinal and categoric data. Text is often processed to be categorized and measured, e.g. word counts, sentiment, emotions, topics, attention, salience and so on.

However, it is well-known that measured data has limits: *"Not everything that can be counted counts, and not everything that counts can be counted."* – William Cameron [1]. For example, the Harvard Business Review (HBR) interviewed 4000 executives asking what success means to them and subjective factors were frequent in answers, such as rewarding relationships, and making a difference; as opposed to objective metrics, such as salary and awards [2].

Where objective data is unavailable, subjective measures, can be extracted, e.g. from textual data from interviews. However, subjective measures may have systematic bias; may be uncorrelated with independent objective measures related to the variable of interest; and may be difficult to aggregate [3]. For example, Nielsen Norman found only 0.53 correlation between user satisfaction and user performance across 298 site designs [4], indicating some mismatches between the measures, e.g. some designs that performed better were unsatisfying (or, designs that performed worse were more satisfying).

Furthermore, the author has been involved in various capital markets projects measuring fuzzy concepts. For example, in risk management, financial risk exposure to stock prices or bankruptcies can be quantitatively modelled based on detailed historical data; but cybersecurity risk and legal risk are rapidly evolving with poor historical data available to model and judge financial risk.

Environmental, Social and Governance (ESG) financial metrics are even more challenging, as it is difficult to define beyond easy-to-measure topics such as carbon emissions or gender/racial demographics of employees [5, 6].

In the author's discussions with a senior executive managing the planning of critical infrastructure, quantitative modeling was described as more well defined and easier to do, compared to qualitative aspects such as community engagement, environmental reviews, and political approvals.

Thus, the position of this paper is visualization focusing only on the measurable may miss critical information. Therefore, it is necessary to have visualizations that include hard-to-measure textual passages. Prior examples are shown in Section 2, including cultural maps, a technique to collect and depict qualitative statements. Section 3 contributes two major approaches to defining spaces within visualizations (area marks/implied space; prose layout); and then use of NLP approaches to fit text into the space.

## 2 APPROACHES TO VISUALIZE NON-QUANTITATIVE

Many prior visualizations attempt to address the non-quantitative from a few different approaches:

### 2.1 Uncertainty visualization

From the perspective of information visualization, there are approaches to deal with fuzziness. One approach is to also model and quantify uncertainty and then provide visual representations of uncertainty, with error bars best known, but also encodings such as blur, contours, de-saturation, and so on, e.g. [7].

### 2.2 Visualization with labels

Within the visualization community, it may be popular to use visualizations with labels of one or two words. The most popular example is the word cloud (e.g. [8]); although there are many other label-centric visualizations such as labeled scatterplots; maps with many labels; or unit visualizations made of words stacked, listed or clustered. NLP may have a role in these visualizations, for example, to extract the words used in unit visualizations. However, these visualizations do not go deeply into the unmeasurable – the semantics of sentences and paragraphs are lost when isolated words are extracted and depicted without context.

### 2.3 Visualization combined with subjective data

Another approach is to visualize the subjective data in reference to quantitative data. *"Cultural mapping is used in both a literal and a metaphorical senses, where it goes beyond strict cartography to include other cultural resources than land, anthropological, sociological, archaeological, genealogical, linguistic, topographical, musicological and botanical, which would be recorded by other appropriate techniques and equipment than maps."* – Peter Poole [9].

Cultural mapping examples may include maps, text, photos, video. The text and the visualizations may be separate and cross-referenced; or combined, such as the examples shown in Figure 1. The left column in Figure 1 shows cartographic maps, with associated annotations in 3D (top) [10], collaborative sticky notes in a workshop (center) [11], and a room sized, heavily annotated cultural map in a museum (bottom) [12]. Cultural maps can be more varied as shown in the right column of Figure 1, with a typographic map (top) [13], annotations combined in a Google Street View mashup (middle) [14], and a relationship graph (bottom) [15].

---

* rbrath@unchartedsoftware.com

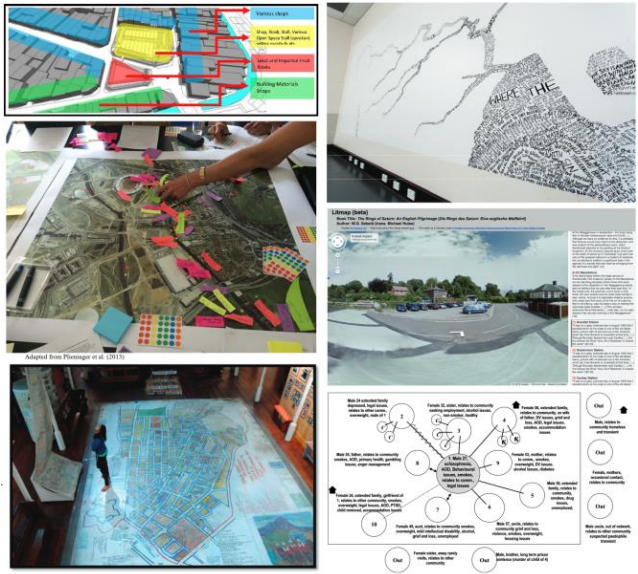

Figure 1: A variety of cultural maps, each with associated prose.

Similar text-rich information graphics with sentences are used by artists and designers as well, such as Kate McLean's sensory maps with associated text (as well as graphs and flowcharts) as well as many prior examples cited in her thesis [16].

## 2.4 Information visualization of running text

Within the visualization community, early examples of visualization with running text can be seen from MIT Media Lab and Muriel Cooper's Visual Language Workshop such as Small's *Talmud* [17] or Fry's *Tendril* [18]: both showing extensive text passages in 3D interactive visualizations (with some occlusion).

Another approach with running text is to encode via text size (e.g. *TextViewer* [19], *Distorted Thumbnails* [20], *Semantize* [21]), or to align running text based on common stems (*WordTree*[22]). Alternatively, running text may be kept consistently sized to facilitate reading and then highlighted by visual encodings such as text color, background highlight, outlines, linkages, bold, italic, geometric shapes and so forth (e.g. *Inkblots* [23], *FeatureInsight* [24], *Poemage* [25], and text highlighting techniques [26]).

*Varifocal Reader* provides multi-level navigation of books from chapter headings down to raw text with intermediate levels such as extracted topics and keywords [27].

Like Tendril, the text knots of Elli et al. [28], has portions of readable and unreadable text fragments, the text in the knots encoding statements of sexual harassment, detangled on interactive engagement.

## 3 EMBEDDING PASSAGES IN VISUALIZATION

Going beyond labels requires representations with suitable space to fit longer passages of text, either within spaces inside a quantitative visualization (such as most of the cultural mapping examples) in Section 3.1; or in coordination with running text (in Section 3.2).

## 3.1 Summarized text in quantitative marks

Innovative typographic cultural maps show the use of large areas to embed significant textual passages directly within the area (Figure 1 top right); varying the orientation of sentences and font sizes to convey additional semantics beyond the textual content and boundaries of area marks.

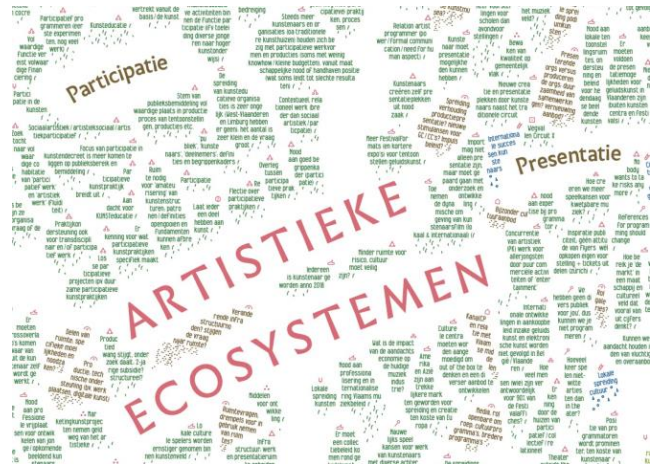

Figure 2: Hierarchical Voronoi treemap of text regarding trends in the arts.

Similarly, the cultural typographic map in Figure 2 (from Speculoos [29]) is a non-geographic map of trends in the professional Flemish arts. It has areas filled with text in a layout similar to a Voronoi tree-map visualization, with text at the lowest level (individual statements) as well as a hierarchy of headings.

### 3.1.1 Text in Area Marks

Generalizing from cultural mapping examples of text in areas on maps and Voronoi treemaps, running text can be placed inside area-based marks in visualizations. Area-based marks are used in visualizations such as choropleth maps, cartograms, streamgraphs, icicle plots, circle packing, hexbins, and treemaps.

For example, in Figure 3, a treemap attempts to show a relationship between oil exporting nations and human rights abuses. Oil exports are indicated with the primary encoding, size, e.g. Saudi Arabia is the largest exporter, Russia is second largest. Color indicates a second metric: *Global Peace Index*. This metric indicates that Russia is worse than Saudi Arabia, but there are worse countries such as Iraq, Libya, South Sudan, and Congo.

But what is a peace index? The Global Peace Index is an amalgam of social, political, and economic factors. It is not obvious what the components are and their respective weights without some form of drill-down; and then the measures are still somewhat subjective comparing not entirely equivalent attributes between countries.

Instead, passages of text can be directly depicted within the treemap areas, in this example, a set of descriptive sentences per country from *Human Rights Watch* (HRW). The text can be directly read: Saudi Arabia's human rights record indicate issues with official accountability for the murder of Jamal Khashoggi; Russia's record indicates it is the most repressive since the Soviet era; UAE detains dissidents even after completing their sentences (and UAE is positively biased on the peace index). Even large exporting countries with generally good peace records, such as Canada and USA, have enough space to indicate rights issues such as the rights of indigenous peoples in Canada, or poverty and inequality in USA.

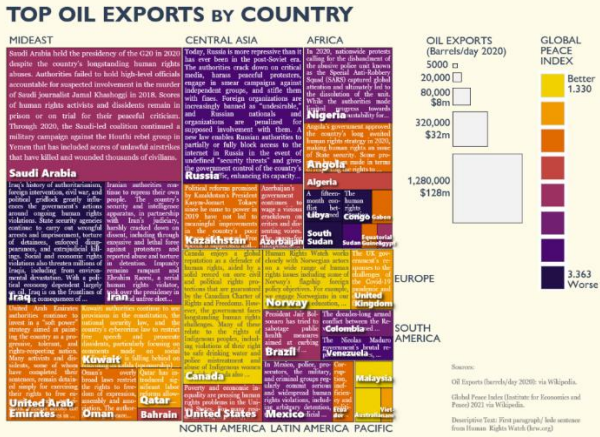

Figure 3: Treemap indicating top oil exporting nations (size), each nations' *Global Peace Index* (color), and corresponding lede text from *Human Rights Watch*.

The different kinds of rights issues not visible with a singular metric can become directly visible with the addition of these extended annotations. Income inequality and indigenous issues are human rights issues as are other repressions, but the viewer can make a more informed comparison about the instances, breadth, severity and cruelty of the largest exporters. Abstract concepts such as peace and corruption are made more concrete with examples.

One may note that larger exporters have larger size and thus more text. Arguably, the largest exporters should have more scrutiny, which in turn is facilitated by more text.

Where does NLP fit in? Finding and fitting an appropriate subset of text into defined areas may be challenging. In the above example, only a fragment of HRW's text fits per cell. Using contemporary NLP large language models (LLMs), such as BERT or GPT (e.g. [30, 31]), can facilitate summarizing text down to a target number of words. In HRW's case, the journalistic writing style uses an inverse pyramid story structure which places the most important information first, thereby making extraction (and truncation) of the lede sentence effective for annotating the visualization.

There is opportunity for additional textual refinement. Some areas are smaller allowing for fewer characters to fit. NLP can further assist:

- A. *Synonyms*. For tight spaces, short word synonyms can be substituted for longer words, e.g. "exodus" or "migrate" instead "displacement"
- B. *Keywords*. In Figure 3, the area for Libya only fits five words and Congo only fits three words. An LLM could extract key phrases or words from a long summary sentence, e.g. *"The human rights situation across the Democratic Republic of Congo remains dire, with internal conflicts and poor governance contributing to a severe food crisis and the internal displacement of nearly 5.5 million people, more than anywhere else in Africa,"* perhaps selecting words such as *"food crisis displacement."*

### 3.1.2 Text in a mark's implied space

Visualization layouts that organize marks into columns, or rows, or circular segments imply space beyond the extents of the mark. For example, within a bar chart, the empty space beyond the top of a bar is allocated to that bar. Thus, any text above the bar is associated with the bar. This space above the bar is often used in bar charts to simply label the bar. However, the entire column is available for text and this space is readily associated with the bar. Figure 4 shows a simple bar chart, where the bars show comparable quantitative metrics for bar heights, and the overlay text indicates qualitative text regarding each bar. In this example, the text is from a report comparing cloud services vendors. The paragraphs relevant to each vendor were summarized with NLP into a couple of sentences, then plotted over top the relevant column.

The approach of using both the positive and negative space associated with the mark could be also applied to distributions, pie charts, hierarchical pie charts, chord diagrams, and so on.

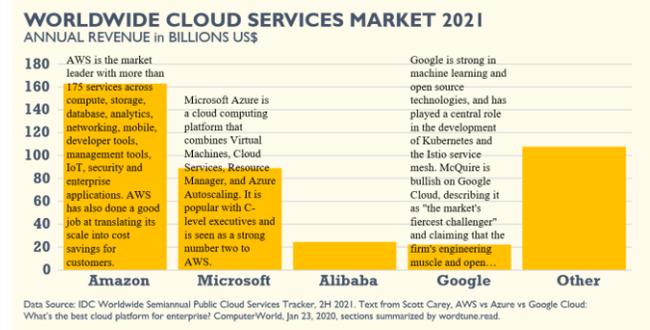

Figure 4: Cloud revenue by vendor, with text indicating vendors' strengths.

## 3.2 Visualizing prose and summaries

Visualization techniques can be applied to prose, either by adding quantitative and categoric data to words in the prose; or by adding or summarizing qualitative data, e.g. on interaction.

### 3.2.1 Markup with quantitative or categoric data

Many of the examples in section 2.D embed quantitative data into running text. More generally, Brath [32] discusses SparkWords as a means of embedding data into running text. In NLP, this approach is extensively used to markup text as seen in a variety of NLP explainability visualizations, such as word salience (e.g. Figure 5 [33]), or word probabilities (e.g. [34]). The markup aids a model developer to better understand what their model is attending to and whether the attention matches their expectation. For example, in Figure 5 a sentiment model is mostly attending to words *stiff* and *lifeless*, which would be expected by the modeler.

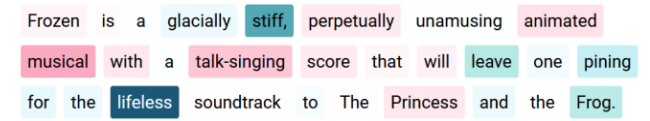

Figure 5: NLP text sentiment task, showing which words are being attended to by the model more (green) and less (red).

The same approach can be applied to summarization, to indicate the source text most attended to, as shown in Figure 6, with source text left, summarized text right, and attended text in the source text highlighted.

Figure 6: Summarization (right) with attended fragments in the source highlighted (left).

Figure 7: Third chapter of *Alice in Wonderland,* with passages summarized into sentences; and interactions such as pop-ups and expansion to show full text.

### 3.2.2 Visual summarization and interaction

NLP and visualization have much potential to enhance tasks associated with prose beyond understanding the model. Literature, legal documents, news, reports, conversations, music lyrics, and so forth are all sources with tasks such as reading and comprehension which may require moving back-and-forth within the text, cross-referencing across passages, navigating through documents, and so on. These tasks are not about quantified data.

Summarization of books is an active area of LLM research. For example, Wu et al. [35], summarize entire books by recursively summarizing sections of text, but their primary use is only with the summarized text and not interacting through the levels to read the underlying prose.

The oft-cited information seeking mantra [36], "overview first, zoom and filter, details on demand," is a general pattern that can be applied to textual documents. For example, the entirety of a novel cannot have an overview, other than using a very coarse proxy of chapter titles. Viewers of online books on web and mobile have few landmarks to aid navigation, except perhaps interactions back to the table of contents and start of chapters. Varifocal Reader [27] solves this problem by simultaneously displaying and linking multiple levels of a book (chapter, topic/word cloud, text) but requires a large display to simultaneous show all levels side-by-side. Readers of a book on mobile must contend with a lot of scrolling within a chapter to refer to an earlier passage. Mobile users do not have physical cues and spatial memory available to readers of a physical book, who can use memory cues such as recollection of approximate position of text on a page or approximate number of pages back.

Figure 8: First three paragraphs of *Alice in Wonderland,* with terse summarizations in a large font under the prose as landmarks and full prose for close reading.

Instead, LLM summarization can turn each few paragraphs into a sentence, such that a chapter fits within a screen or two. Then the reader can use interactions to access full text, such as zoom, tooltips, etc. Figure 7 shows the third chapter of *Alice's Adventures in Wonderland*, where a LLM (cohere.ai large 202206 model) has been used to summarize each few paragraphs (50-120 words) into a sentence. All the sentences for a chapter fit within a mobile view. Interactions include tap and hold popup (i.e. tooltip) shown in the black bubble; or individual sentences tapped and expanded in place to show the full contents (at a slightly smaller font size). An information hierarchy is applied with largest fonts for the formal headings (i.e. chapter title); slightly larger text for passage summaries, and slightly smaller text for full text. (See https://codepen.io/Rbrath/full/ZEoBdZQ for an interactive demonstration).

Another alternative is to superimpose the detailed text over the summarized text, thereby creating large landmark text within the full prose [32], as per the example in Figure 8 showing the first few paragraphs of *Wonderland*. This provides a means to rapidly skim the larger summary text, then focus to attend the detail text. However, text-on-text has issues with occlusion and legibility, presumably slowing reading speed, and does not reduce the space.

## 4 DISCUSSION

NLP and specifically LLM summarization create a new opportunity to reduce text to fit within limited areas in a visualization while retaining most of the semantics of the text. Text is not always well structured, but the performance and quality of LLM's is rapidly advancing with remarkable capabilities beyond summarization. For example, Google PaLM with 500 billion parameters can explain jokes and perform logical inference [37]. Instead of summarization, future versions of this approach might use NLP to explain quantitative values with one-shot prompts (e.g. prompt: "Microsoft Azure is popular because:", DeepAI response: "It enables enterprises to easily manage database data from Oracle, Azure AD, Google Cloud Platform and enterprise systems with increased performance. Because it integrates with most of the current Microsoft Azure applications on their servers using Azure DBAM and provides a high level of availability and performance across multiple devices and platforms.")

Note that LLM summarization does not generate consistent lengths of text, as shown in Figure 4 or 7. This is an open question for steering LLM's via prompting or fine-tuning. Autoregressive LLM's (e.g. GPT) generate a sequence of incremental words based only on the prior words. Autoregressive LLM's may be difficult to use to generate text within a fixed target number of words or characters. The author hypothesizes that bi-directional LLM's, such as BERT, which are trained on both preceding and trailing word contexts, may perform better given a constraint with a target number of words or characters.

The examples here are simple with minimal markup. Additional visualization attributes, for categoric and quantitative data such as text color, font-weight, italics, background color and so forth can be used to indicate additional data of interest. For example, in Figure 7, characters could be bolded (e.g. Mouse, Lory, Alice, White Rabbit, etc.); or the LLM can be used to identify the most semantically important sentences (e.g. via NLP extractive summarization); or perhaps italicizing sentences based on NLP metrics such as confidence or similarity to aid human-in-the-loop editing (e.g. Alice does not save the White Rabbit's life, nor catch a bird). Additional text visualization techniques can be applied to this summarized text. For example, each line is a sentence, and many sentences repeat words. For example, characters such as Alice, the Mouse, the Lory, and the Dodo recur frequently in Chapter 3, suggesting potential alternative layouts such as WordTree [22] to indicate the sections where Alice is the first word in successive sentences; or possibly a means of indicating character relationships using a variant of story timelines [e.g. 37, based on 38] or kelp fusion [39].

## 5 CONCLUSION

Large language models completely revolutionize the potential to fit relevant text within the space constraints of visualizations. This quick paper only hints at the possibilities by attempting to frame some variations using different NLP approaches and different visualization layouts.

Images by the author (Figures 3, 4, 7 and 8) are licensed under CC-BY-SA-4.0.